# Can impact excitation explain efficient carrier multiplication in carbon nanotube photodiodes?

### Roi Baer[a] and Eran Rabani[b]

*(a) Fritz Haber Center for Molecular Dynamics, the Chaim Weizmann Institute of Chemistry, the Hebrew University of Jerusalem, Jerusalem 91904 Israel.(b) School of Chemistry, The Raymond and Beverly Sackler Faculty of Exact Sciences, Tel Aviv University, Tel Aviv 69978 Israel*

We address recent experiments (Science 325, 1367 (2009)) reporting on highly efficient multiplication of electron-hole pairs in carbon nanotube photodiodes at photon energies near the carrier multiplication threshold (twice the quasi-particle band gap). This result is surprising in light of recent experimental and theoretical work on multiexciton generation in other confined materials, such as semiconducting nanocrystals. We propose a detailed mechanism based on carrier dynamics and impact excitation resulting in highly efficient multiplication of electron-hole pairs. We discuss the important time and energy scales of the problem and provide analysis of the role of temperature and the length of the diode.

Carrier multiplication (CM) is a process where several charge carriers are generated upon the absorption of a single photon in semiconductors.[1] This process is of great technological ramifications for solar cells and other light harvesting technologies.[2] For example, it is expected that if more charge carriers are created shortly after the photon is absorbed, a larger fraction of the photon energy can successfully be converted into electricity, thus increasing device efficiency.

Recent experiments by Gabor *et al* [3] reported highly efficient generation of electron-hole (e-h) pairs in single-walled carbon nanotube (CNT) p-n junction photodiodes. Specifically, once the exciting optical photon energy exceeds twice the quasi-particle band gap ($E_g$), a photocurrent step is observed in the I-V source-drain characteristics of the device. The steps were associated with the efficient generation of additional e-h pairs by impact excitation once the photon has enough energy to populate the $\varepsilon_{e2}$ - $\varepsilon_{h2}$ bands (see Figure 1a).

This result is important and surprising in light of the controversy concerning charge multiplication in other low-dimensional system, such as semiconducting nanocrystals (NCs).[4] Recent studies[5,6] have questioned the efficiency of charge multiplication in NCs at photon energies below $3E_g$. Thus, understanding the mechanism by which e-h pair generation is efficient is CNTs is of great interest.

In this letter, we present a theoretical framework that provides new insights into this problem. We consider a zigzag CNT under the experimental conditions of Ref. 3 as depicted in Figure 1b. The CNT is subjected to an external split gate voltage with $V_1 = -V_2$ inducing an electric field inside the CNT along its axis. As a result, a p-n junction is formed with a band structure shown schematically in Figure 1b (shown are the bottom of the conduction and top of the valence bands). As described by Gabor et al,[3] the devices shows ordinary p-n junction transport characteristics in the absence of light (very small reverse bias current and a threshold voltage for turn-on in the forward bias). When illuminated at low photon energies above $E_g$ and below $2E_g$, the device shows constant photocurrent at reverse bias, intersecting zero at the open-circuit voltage, $eV_{OC} \approx E_g$, typical of p-n junctions.[3]

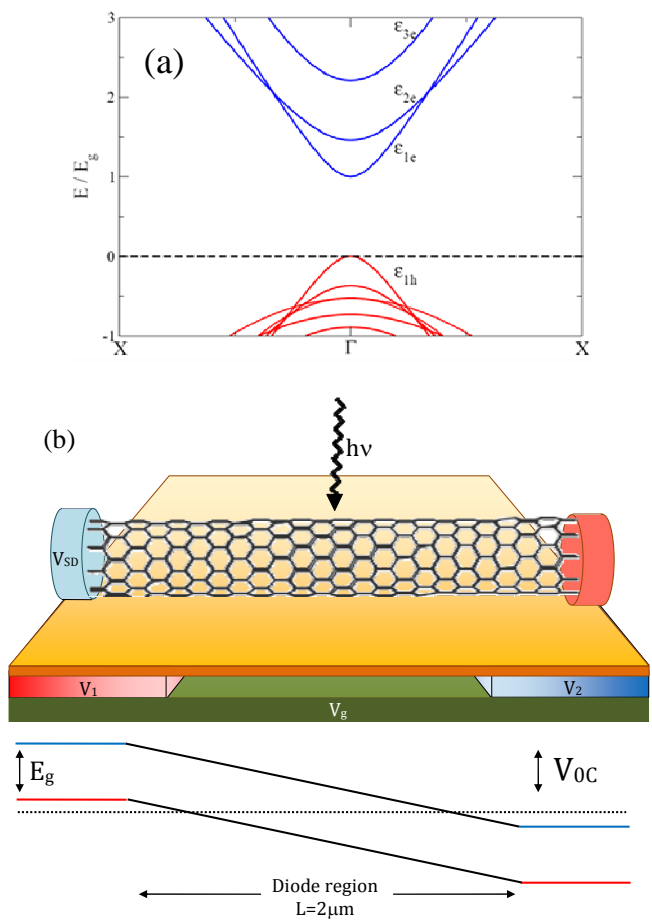

Figure 1: (a) The band structure of a (19,0) zigzag CNT calculated by the extended Hückel method. Bands are labeled as $\varepsilon_{ej}$ and $\varepsilon_{hj}$ for electrons and holes in band *j* respectively. Note that excitation in excess of $2.2E_g$ can create e-h pairs in either $\varepsilon_{e2} - \varepsilon_{h2}$ or $\varepsilon_{3e} - \varepsilon_{1h}$ bands. (b) A sketch of the experimental setup[3] along with the bottom of the conduction and top of the valence energies. The dotted line is the Fermi level. $V_{OC} \approx E_g$ is the open circuit voltage; $V_1 = -V_2$ are the split gate voltages; $V_g$ is overall gate potential; $V_{SD}$ is the source-drain bias.

At photon energies above $2E_g$ a series of steps in the reverse bias photocurrent is observed.[3] These steps were associated with efficient generation of e-h pairs by impact excitation from the $\varepsilon_{e2}$ - $\varepsilon_{h2}$ bands (see Figure 1a) with an onset temperature of about $90°K$.[3] The source-drain bias at which the steps occur depends also on the position of CNT illumination. Gabor *et al*[3] proposed a schematic mechanism by which charges are accelerated by the internal diode field gaining energy in the $\varepsilon_{e2}$ - $\varepsilon_{h2}$ bands and generating multiple charge carriers by impact excitation.

This mechanism provides the seeds for the present work. In addition, several other important aspects need to be considered. First, can impact excitation occur from the $\varepsilon_{e2}$ - $\varepsilon_{h2}$ bands,[7] or should other bands play a role? Second, what is the role of other competing relaxation processes, such as electron-phonon couplings?[8, 9] Third, what is the mechanism by which excitons break into e-h pairs, in particular in light of recent studies implying large exciton binding energy in semiconducting CNTs.[10] We propose the following mechanism for the e-h generation in CNT photodiodes which addresses these issues in some detail (see illustration in Figure 2):

1) Photoexcitation occurs into a quasi-bound excitonic state at energies above $2E_g$.

2) The quasi-bound exciton then breaks into an e-h pair, with an electron in the $\varepsilon_{e3}$ band and a hole in the $\varepsilon_{h1}$ band (alternatively, the exciton may break into an e-h pair with an electron in the lowest band and a hole in an excited band).

3) The free carriers gain energy by accelerating in the diode field, a process we describe by a Langevin equation with phonon induced friction. The maximal carrier energy depends on the ratio between the diode field and phonon friction coefficient.

4) Highly efficient carrier multiplication by impact excitation occurs due to acceleration of carriers to high enough energy, as described in the previous step. All resultant carriers form in lower bands conserving total energy, linear and angular momenta.

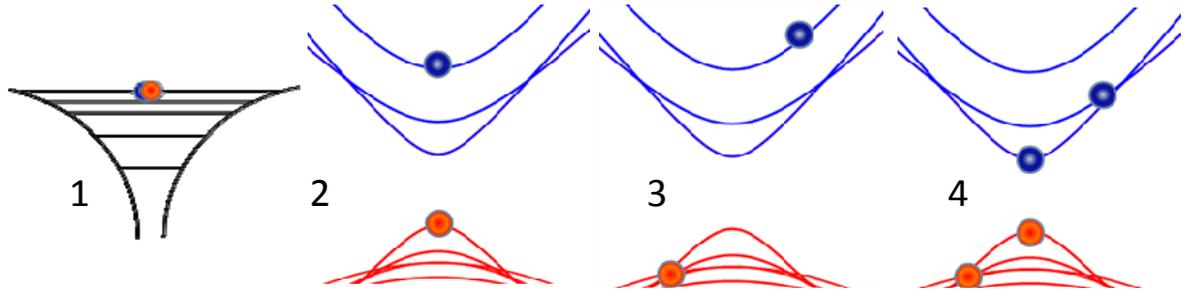

Figure 2: Schematic representation of the four step carrier multiplication by impact excitation mechanism in a CNT photodiode. 1 - photoexcitation into a quasi-bound state. 2 - Exciton breaks into an electron-hole pair in the $\varepsilon_{e3}$ - $\varepsilon_{h1}$ bands. 3 - Energy gain of the carriers by interplay of diode field and phonon emission. 4 – Carrier multiplication by impact excitation.

The first step involves excitation into a *quasi-bound* exciton. The observed photocurrent at zero source-drain bias supports the claim that this happens for photon energies is in excess of twice the quasiparticle band-gap ($2E_g$). If excitation created only a bound e-h pair (exciton) photocurrent would have been significantly suppressed at zero bias. It should be noted that the internal diode field is not strong enough to break a bound exciton. This is because the exciton binding energy is $E_B = \frac{e^2}{\varepsilon a_B} \approx \frac{1}{3} E_g$ where $a_B = \frac{\hbar^2 \varepsilon}{e^2 \mu^*} \approx 3\text{nm}$ is the exciton Bohr radius and $\varepsilon = 3.3$ is the relevant dielectric constant,[10] while the internal field across the exciton contributes only a negligible perturbation, $eV_{OC} \frac{a_B}{L} \approx 10^{-3} E_g$.

In the second step, the quasi-bound exciton decays into a resonant e-h pair state. In the energy range of interest, the exciton can spontaneously dissociate into e-h pair either in the $\varepsilon_{e2}$ - $\varepsilon_{h2}$ bands or in the $\varepsilon_{e3}$ - $\varepsilon_{h1}$ bands (see Figure 1a). For the sake of simplicity, from now on, we ignore the process by which the exciton dissociates into an electron in the ground electronic band ($\varepsilon_{e1}$) and the hole into some other excited valence band which is also in resonant with the $\varepsilon_{e2}$ - $\varepsilon_{h2}$ energy. However, the proposed mechanism applies for this case as well. Below we argue that to observe carrier multiplication by impact excitation, the excitation process *must* end up with an e-h pair in the $\varepsilon_{e3}$ - $\varepsilon_{h1}$ bands. In fact, this is consistent with the experimental observation that only one charge multiplication occurs upon excitation at slightly above $2E_g$ rather than two carrier multiplications events that would occur in the symmetric $\varepsilon_{e2}$ - $\varepsilon_{h2}$ case, if impact excitation would be allowed from these bands. The formation of e-h pairs in the $\varepsilon_{e3}$ - $\varepsilon_{h1}$ bands can occur by direct absorption (depending on light polarization) or by coupling the $\varepsilon_{e2}$ - $\varepsilon_{h2}$ bands with the nearly iso-energetic $\varepsilon_{e3}$ - $\varepsilon_{h1}$ bands by a per-turbation that permits a change in the angular momentum, such as the presence of a high dielectric surface.

In the third step, once separated, each carrier accelerates in its band by the internal diode field in opposite directions and experiences dissipation induced by coupling to acoustic and optical phonons. The longitudinal motion of these quasiparticle carriers is described by a classical Langevin equation:

$$\dot{\nu}(t) = -\frac{\gamma_{ep}(\varepsilon_i)}{\nu}\sqrt{1+\nu^2}\left(\sqrt{1+\nu^2}-1\right) + \frac{F(z)}{m_i^* v_{0,i}}$$
$$\dot{z}(t) = \frac{v_{0,i}}{\sqrt{1+\nu^{-2}}}. \quad (1)$$

To derive Eq.(1), we assumed a Dirac energy dispersion relation[9] $\varepsilon_i(\nu) = m_i^* v_{0,i}^2 \sqrt{1+\nu^2}$, where $m_i^*$ and $v_{0,i}$ are the "rest mass" and the "speed of light" parameters (see Table 1 for values of these parameters to the extended Hückel dispersion relations) and $\nu = \frac{\hbar k}{m_i^* v_{0,i}}$ is a scaled momentum. In the above equation, $F(z) = qV_{OC}\theta(L-z)\theta(z)$ is the force exerted by the internal diode field on the quasi-particle of charge $q = \pm e$, assumed to be uniform along the tube axis ($z$) in the diode region of length $L$ (see Figure 1). For simplicity, we neglect the random fluctuating force in the Langevin description. $\gamma_{ep} = \gamma_{ac} + \gamma_{opt}(\varepsilon)$ is the friction coefficient, describing the irreversible energy loss to phonons. We assume diffusive regime for the transport with $\gamma_{ac} = \alpha\frac{T}{d}$ where[9] $\alpha = 10^{-2}\frac{nm}{K\,ps}$ represents the acoustic phonon friction coefficient; $\gamma_{opt}(\varepsilon) = \gamma_{opt,0}\theta(\varepsilon - \hbar\omega_{opt})$, where $\gamma_{opt,0} = 50\,ps^{-1}$ and $\hbar\omega_{opt} = 0.2\,eV$ represents the phonon friction arising from C-C optical mode of the CNT.[8, 11] We neglect the nonadiabatic inter-band transitions by optical phonons which occur on much longer timescales.[12]

Table 1: Dirac parameters fitted to the extended Hückel band dispersion relations. Also shown the rest energies, corresponding to $E_{ej}$ for electrons and $-E_{hj}$ for holes.

| b a n d | Electrons | | | Holes | | |
|---|---|---|---|---|---|---|
| | $m^*$ $(m_e)$ | $v_0$ $(10^5\,m/s)$ | $m^* v_0^2$ $(E_g)$ | $m^*$ $(m_e)$ | $v_0$ $(10^5\,m/s)$ | $m^* v_0^2$ $(E_g)$ |
| 1 | 0.07 | 7.8 | 0.50 | 0.08 | 7.1 | 0.49 |
| 2 | 0.18 | 6.4 | 0.94 | 0.20 | 5.8 | 0.86 |
| 3 | 0.17 | 8.9 | 1.70 | | | |

The energy $E(t)$ (measured from the bottom of the first conduction band minimum $E_{e1}$) and the scaled momentum $\nu(t)$ of an electron in band $\varepsilon_{3e}$ as a function of time are shown in Figure 3a and 3b, respectively, for various temperatures. The initial position and momentum for all trajectories are $z(0) = L/2$ and $\nu(0) = 0$, corresponding to an excitation in the middle of the tube with photon energies $\varepsilon_{e3} - \varepsilon_{h1}$. The initial electron energy is $E(0) = m_3^* v_{0,3}^2 \sqrt{1+\nu^2(0)} - E_{e1} \approx 1.2 E_g$ (where $E_{ej}$ ($E_{hj}$) is the j$^{th}$-conduction (valance) band minimum (maximum)) increases due to the acceleration of the electron by the diode field, reaching a maximum after about 3-5 ps, when the electron leaves the region of the diode. The total energy gained is smaller than the maximal value of $eV_{OC}/2 \approx E_g/2$ as a result of energy loss to acoustic and optical phonons. The maximal energy attained decreases with increasing temperature reflecting the temperature dependence of the acoustic phonon friction. The coupling to optical phonons is activated only after the electron accelerates to energies above 0.2eV (about $\frac{1}{2}E_g$ for the current tube), and thus the impact of optical phonons on the trajectories shown in Figure 3 can be ignored for most temperatures, since the energy trajectory remains below this value. The increase in the carrier's energy is crucial for the carrier multiplication, since only at sufficiently high energies charge multiplication becomes efficient, as will be shown below. The efficiency of charge multiplication will depend on the temperature reflecting the behavior shown in Figure 3a.

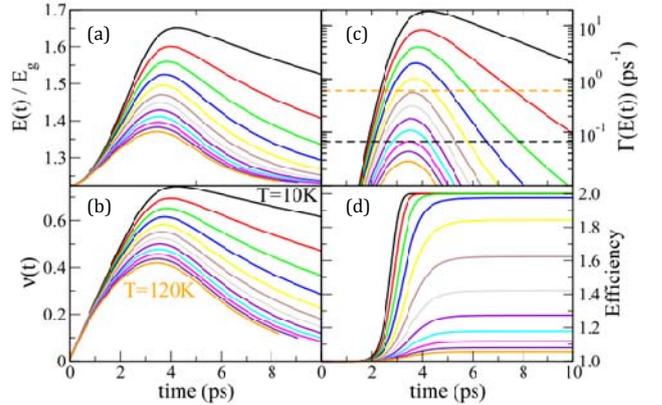

Figure 3: The energy above the conduction band minimum $E(t) = m_3^* v_{0,3}^2 \sqrt{1+\nu(t)^2} - E_{e1}$ (a), the scaled momentum $\nu(t)$ (b), the trion generation rates (c) and efficiencies (d) as a function of time for several temperatures in a (19,0) zigzag CNT of $d = 1.5\,nm$ (19,0), $L = 2\,\mu m$ and $eV_{OC} = E_g = 0.44\,eV$. The dashed lines in panel (c) represent the acoustic friction parameter at $T = 10\,^oK$ (black) and $T = 120\,^oK$ (orange). Sequential curves represent temperature increments of $10\,^oK$.

The last step of the proposed mechanism involves charge multiplication by impact excitation of the hot accelerated electrons. We describe this process by considering the free negative charge carrier population in the $\varepsilon_{3e}$ band at time $t$ having energy $\mathcal{E}$, $S(\mathcal{E},t)d\mathcal{E}$, ($\mathcal{E}$ measures the energy above the conduction band minimum or, for holes, below the valence band maximum). We assume that $S(\mathcal{E},t)$ obeys a drift equation with a sink term, following the trajectory $E(t) = m_3^* v_{0,3}^2 \sqrt{1+\nu(t)^2} - E_{e1}$ obtained from the Langevin equation (Eq. (1)):

$$\frac{\partial S(\mathcal{E},t)}{\partial t} = -\dot{E}(t)\frac{\partial S(\mathcal{E},t)}{\partial \mathcal{E}} - \Gamma(\mathcal{E})S(\mathcal{E},t) \qquad (2)$$

The initial condition is $S(\mathcal{E},0) = \delta(\mathcal{E}-E(0))$ describing a single carrier at the bottom of the $\varepsilon_{e3}$ band. Defining $G(t) = \int_0^t \Gamma(E(\tau))d\tau$, the solution of Eq. (2) is $S(\mathcal{E},t) = \delta(\mathcal{E}-E(t))e^{-G(t)}$. Integrating over $\mathcal{E}$, one finds that the photocurrent generation efficiency $\eta(t)$ is given by:

$$\eta(t) = 2 - e^{-G(t)} \qquad (3)$$

In Eq.(2), $\Gamma(\mathcal{E})S(\mathcal{E},t)$ is the extinction rate of negative charge carriers from the $\varepsilon_{3e}$ band to a negative trion (two electrons and one hole).[6] $\Gamma(\mathcal{E})$ is the energy dependent rate constant of trion generation assumed of the ad-hoc form:

$$\frac{\Gamma(\mathcal{E})}{\Gamma_\infty} = \left(\frac{\Gamma_g}{\Gamma_\infty}\right)^{\exp\{-\kappa(\mathcal{E}/E_g - 1)\}} \qquad (4)$$

The above expression is valid for $\mathcal{E} > E_g$, otherwise the rate is zero. The three parameters $\Gamma_\infty$, $\Gamma_g$ and $\kappa$ are determined by a fit to the trion generation rates calculated for all electrons of energy $\mathcal{E}$ using the Fermi Golden rule (for a full derivation see Ref. 6):

$$\Gamma_a = \frac{4\pi}{\hbar}\sum_{cbj}\left|\left(2V_{acjb} - V_{abjc}\right)\right|^2 \delta\left(\varepsilon_a - (\varepsilon_b + \varepsilon_c - \varepsilon_j)\right) \qquad (5)$$

where the indexes $a,b$ and $c$ correspond to electron levels and $j$ to a hole level. The calculation is preformed in real space with a converging super-cell. The two-electron integrals are:[13]

$$V_{abjc} = \int \psi_a(\mathbf{r}_2)\psi_b(\mathbf{r}_2)\phi_{jc}(\mathbf{r}_2)d^3r_2 \qquad (6)$$

Where $\phi_{jc}(\mathbf{r}_2)$ is the "electrostatic potential" at $\mathbf{r}_2$ due to a charge distribution $\psi_j(\mathbf{r}_1)\psi_c(\mathbf{r}_1)$:

$$\phi_{jc}(\mathbf{r}_2) = \frac{1}{\sqrt{\varepsilon_\perp^2 \varepsilon_\parallel}}\int \frac{\psi_j(\mathbf{r}_1)\psi_c(\mathbf{r}_1)}{\sqrt{x_{12}^2/\varepsilon_\perp + y_{12}^2/\varepsilon_\perp + z_{12}^2/\varepsilon_\parallel}}d^3r_1 \qquad (7)$$

Where $\varepsilon_\perp = 2$ and $\varepsilon_\parallel = 6$ are the transverse ($x-y$) and longitudinal ($z$) static permittivity of semiconducting CNTs. These values were chosen to reproduce the exciton binding energy in CNTs.[14] The Coulomb matrix elements $V_{abjc}$ were calculated by representing the extended Hückel molecular eigenstates ($\psi_j(\mathbf{r}_1)$) on a 3D real-space grid and applying fast Fourier techniques for the convolution operation in Eq. (7).

The calculated trion generation rates are shown in Figure 4, for several zigzag CNTs with varying diameters, as a function of the scaled energy $\mathcal{E}/E_g$. Each point represents an initial electron (decaying to a negative trion) or hole (decaying to a positive trion) excitation. The density of holes (the number of points) is larger than the density of electron, as clearly evident in the figure. The rates are zero for $\mathcal{E} < E_g$ implying that electrons or holes in the second band $\varepsilon_{e2}$ ($E_{e2} - E_{e1} \approx \frac{1}{2}E_g$) or $\varepsilon_{h2}$ ($E_{h1} - E_{h2} \approx \frac{1}{2}E_g$) cannot lead to the formation of trions, consistent with earlier calculation within a simpler electronic structure approach.[7]

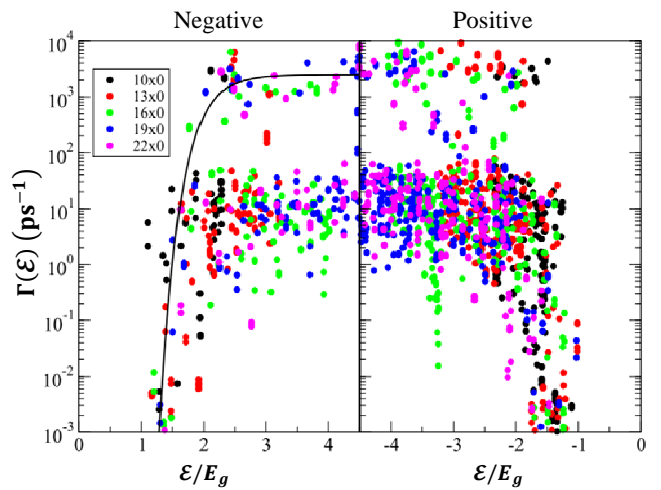

Figure 4: The negative (positive) trion generation rates as a function of energy above (below) the conduction (valence) band minimum (maximum) for several zigzag CNTs. The solid line are a fit of the maximal negative trion rates to Eq. (4) with parameters: $\Gamma_\infty = 2.5\,\text{fs}^{-1}$, $\Gamma_0 = 10^{-15}\Gamma_\infty$, $\kappa = 3$.

For electrons or holes, the rates span a wide range of values (several orders of magnitude), increasing rapidly as the carrier energy exceeds $1.2E_g$ and reaching a plateau for energies above $2E_g$. The magnitude of the plateau rate is relatively insensitive to the diameter $d$ of the CNTs despite the fact that $E_g$ is inversely proportional to $d$.[15]

The plateau rate is characterized by two distinct populations probably resulting from the complex band structure at high energies where several bands overlap. We fit the parameters in Eq. (4) to the maximal values of the rates and not to the average value. The diode field mixes the different states (the equivalent Schrödinger picture for charge acceleration) and thus, the decay to a trion occurs from states with the largest decay rate. We note that for the parameters used in Figure 3, it is the rapid increase of the $\Gamma(\varepsilon)$ rather than the plateau value $\Gamma_\infty$ that affects the efficiency of the carrier multiplication process. Even for smaller values of $\Gamma_\infty$, such as an average rate value, the picture described in Figure 3 is qualitatively preserved.

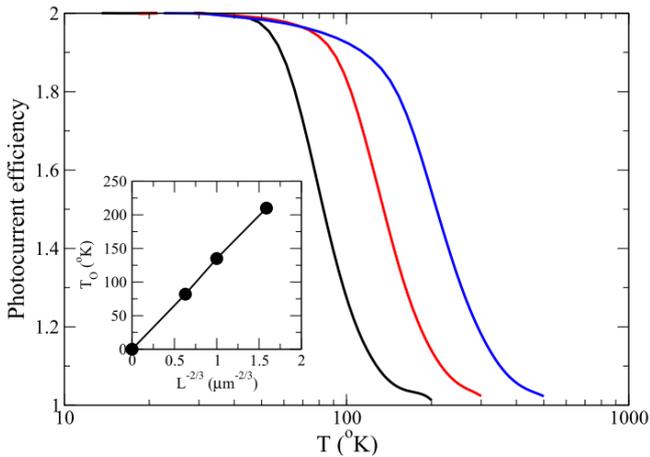

Figure 5: Photocurrent efficiency as a function of temperature for $L = \frac{1}{2}\mu$m (blue curve), $L = 1\mu$m (red curve), and $L = 2\mu$m (black curve). Inset: Onset temperature versus $L^{-2/3}$ confirming the scaling law.

Returning to the drift equation (Eq. (2)), we show in Figure 3c the trion generation rate $\Gamma(E(t))$, as a function of time for various temperatures. The rates follow the time evolution of the energy trajectory, initially increasing, reaching a maximum and then decreasing. The value of the maximum of $\Gamma(E(t))$ in Figure 3c should be compared to the phonon induced friction coefficient $\gamma_{ph}$. At low temperatures it rises significantly above $\gamma_{ph}$ and the photocurrent efficiency $\eta$ approaches its maximal value of 2, as shown in Figure 3d. As temperature increases the maximal trion generation rate drops below $\gamma_{ph}$ and $\eta$ cannot significantly exceed the value of 1. Therefore, there is clearly an onset temperature, $T_O$, below which charge multiplication leading to an increase in photocurrent is possible.

In Figure 5 we show the temperature dependence of the total efficiency $\eta_\infty = \eta(t \to \infty)$ for various diode field lengths ($L$). The switching behavior is qualitatively similar to the temperature dependence observed by Gabor et al[3]. We define an onset temperature $T_O$ as the temperature at which $\eta_\infty = 1\frac{1}{2}$. We find that it scales as $L^{-2/3}$, as shown in the inset.

In summary, we proposed a mechanism based on impact excitation which accounts for highly efficient multiple e-h pair generation in CNT photodiodes. The mechanism involves photoexcitation into a quasi-bound exciton which breaks into an e-h pair in the $\varepsilon_{e3} - \varepsilon_{h1}$ bands. The excited charge carriers are then accelerated by the diode field under phonon induced friction. The carrier dynamics were described by a Langevin equation where the interplay between the field and the rate by phonon decay (which is temperature dependent) sets the maximal energy reached by the charge carriers. At low frictions (low temperature), the charges are accelerated to high enough energies where charge carrier multiplication is highly efficient. The photocurrent efficiency was described by a drift equation with a sink term, where the rate of charge multiplication was calculated within a Fermi Golden rule within an extended Hückel electronic structure model. Several conclusions can be drawn from our work:

1) Impact excitation (trion formation) can explain highly efficient carrier multiplication in CNTs photodiodes, but only for electrons in the $\varepsilon_{e3}$ band or higher. Equivalently, a similar process can occur for holes excited to bands with high enough energy. Excitation into the $\varepsilon_{e2} - \varepsilon_{h2}$ bands *cannot* lead to carrier multiplication by impact excitation.

2) The proposed mechanism provides an onset energy to observe carrier multiplication slightly above $2E_g$, since below there is not sufficient photon energy to excite the system into the $\varepsilon_{e3} - \varepsilon_{h1}$ band.

3) Excitation into the $\varepsilon_{e3} - \varepsilon_{h1}$ bands suggests that only *one* charge multiplication event can occur at photon energies slightly above $2E_g$, since e-h symmetry is broken, consistent with experiments.[3] Contrary, excitation into the symmetric $\varepsilon_{e2} - \varepsilon_{h2}$ could lead to the generation of *two* multiplication events, one by the electron and one by the hole, if impact excitation would be allowed from the $\varepsilon_{e2} - \varepsilon_{h2}$ bands, giving rise to 300% efficiencies at photon energies of $2E_g$, which is not seen experimentally.

4) Our proposed mechanism also provides an explanation to the absence of cascade of charge carrier multiplication, since multiple carriers are generated at lower bands. Thus, even if these carriers are accelerated by the diode field, energy conservation restricts carrier multiplication by impact excitation from these bands.

5) We predict an onset temperature to observe carrier multiplication that scales as $L^{-2/3}$, which implies that the phenomena can occur at room temperatures for short enough CNTs ($L < 500\,\text{nm}$).

The present work provides a minimalist model for carrier multiplication in CNT photodiodes. Despite its simplicity, it captures the hallmarks of current experimental observations. Future work will focus on assessing prediction made with more elaborate models. The present work also calls for additional experiments to verify the hypothesis and predictions made here (asymmetric excitation, role of substrate, onset temperature and length scaling).

**Acknowledgments:** We thank Louis Brus, Mark Hybertsen and David Reichman for many useful discussions and the Center for Re-Defining Photovoltaic Efficiency Through Molecule Scale Control, an Energy Frontier Research Center funded by the U.S. Department of Energy, Office of Science, Office of Basic Energy Sciences under Award Number DE-SC0001085 for support. This work was supported by the Israel-US Binational Science Foundation (Grants numbers 2008221 and 2008190).